# Antiferroelectric switching inside ferroelastic domain walls

Guangming Lu[1*], Gustau Catalan[2,3*] and Ekhard K. H. Salje[4*]

[1]*School of Environmental and Materials Engineering, Yantai University, Yantai 264005, China*
[2]*Catalan Institute of Nanoscience and Nanotechnology (ICN2), CSIC and BIST, Campus UAB, Bellaterra, Barcelona, 08193 Catalonia*
[3]*ICREA - Institució Catalana de Recerca i Estudis Avançats, Barcelona, Catalonia, 08010*
[4]*Department of Earth Sciences, University of Cambridge, Cambridge CB2 3EQ, UK*

Corresponding author: luguangming1990@ytu.edu.cn; gustau.catalan@icn2.cat; ekhard@esc.cam.ac.uk

## ABSTRACT

Ferroelastic materials (materials with switchable spontaneous strain) often are centrosymmetric, but their domain walls are always polar, as their internal strain gradients cause polarization via flexoelectricity. This polarization is generally not switchable by an external electric field, because reversing the domain wall polarity would require reversing the strain gradient, which in turn would require switching the spontaneous strain of the adjacent domains, destroying the domain wall in the process. However, domain wall polarization can also arise from biquadratic coupling between polar and non-polar order parameters (e.g. octahedral tilts in perovskites). Such coupling is independent of the sign of the polarization and thus allows switching between +P and -P. In this work, we seek to answer the question of whether the polarization of domain walls in ferroelastic perovskites is switchable, as per the symmetric biquadratic term, or non-switchable due to the unipolar flexoelectric bias. Using perovskite calcium titanate ($CaTiO_3$) as a paradigm, molecular dynamics calculations indicate that high electric fields broaden the ferroelastic domain walls, thereby reducing flexoelectricity (as the domain wall strain gradient is inversely proportional to the wall width), eventually enabling switching. The polarization switching, however, is not ferroelectric-like with a simple hysteresis loop, but antiferroelectric-like with a double hysteresis loop. Ferroelastic domain walls thus behave as functional antiferroelectric elements, and also as nucleation points for a bulk phase transition to a polar state.

**INTRODUCTION**

Ferroelectric and/or ferroelastic domain walls (DWs) contain their own functional properties [1-4]. Examples include free-electron gas confinement within DWs [5,6], DW magnetism [7,8], DW conductivity [9-11], and photovoltaic effects [12,13]. Ferroelastic DWs are noteworthy in this context, as their electronic properties, which may include polarization [14-19], superconductivity [20,21], or magnetism [22-25] stand in stark contrast to those of the domains themselves, which are only defined by their mechanically switchable spontaneous strain [26]. Domain walls can also be moved by application of an external field, and their properties are carried alongside. Thus, for example, using stress to change the ferroelastic domain configuration causes changes in the domain wall distribution, and thus in the volume-averaged polarization of a crystal [18]. Such change in polarization is due to the domain wall motion rather than a change in the internal polarization of the wall, however; switching the internal polarization of individual ferroelastic domain walls remains an unfulfilled challenge.

The emergence of DW properties stems from structural deviations inside them with respect to the parent bulk phase [27,28]. Ferroelastic DWs, for example, are polar (and therefore non-centrosymmetric) even when the bulk domains are centrosymmetric [26]. The structural causes of this DW polarization can be multiple, but one that is always present is flexoelectricity: by linking two domains of opposite spontaneous strain, ferroelastic domain walls inherently possess an internal strain gradient roughly equal to the difference in spontaneous strain across the wall divided by its thickness, and thus also a gradient-induced polarization -i.e. flexoelectricity [29,30]. For the archetypal ferroelastic perovskite calcium titanate ($CaTiO_3$, the eponymous Perovskite mineral), the polarization of the twin domain walls has been theoretically predicted and experimentally established [27,31]. However, the hope that twin walls may be used as memory devices by switching their intrinsic dipoles has never been realized. An important difficulty in this respect is that switching requires two or more stable configurations, whereas linear effects such as flexoelectricity (linear coupling between strain gradient and polarization) are unipolar, and therefore not conducive to a symmetrically switchable polarization [32-38].

On the other hand, flexoelectricity is not the only possible source of polarization. Crystals in general, and perovskites in particular, often possess more than one structural order parameter (e.g. octahedral tilts, spontaneous strain, or polarization), and these can couple to each other and their gradients linearly or quadratically depending on symmetry constraints. The biquadratic Houchmandzadeh-Lajzerowicz-Salje term, $\lambda Q_1^2 P^2$ [39] is allowed in any symmetry (including even magnetic symmetry [40]), and therefore it can always be present. Moreover, since the polarization term is squared, +P and -P have the same energy, so the energy cost of switching between them is symmetric.

In many ferroelastic perovskites, octahedral tilts are typically anticorrelated with ferroelectric polarization [41], and this anticorrelation can be described by positive biquadratic coupling between octahedral tilts and polarization, such that when the tilt is the primary order parameter, the positive sign of the coupling requires the polarization to go to zero to minimize the free energy [39,40]. Conversely, when tilts are cancelled (as it happens inside antiferrodistortive antiphase boundaries) the previously suppressed order parameter –the polarization— can emerge [42]. In Morozovska's work [43], the combined effects of flexoelectricity and biquadratic coupling were considered under the framework of Landau-Ginzburg-Devonshire theory, and thought to be responsible for the spontaneous polarization in the vicinity of structural twin walls in the octahedrally tilted phase of otherwise non-ferroelectric perovskites such as $CaTiO_3$, $SrTiO_3$, and $EuTiO_3$. The polarization was calculated for the domain walls of $SrTiO_3$, a material where the spontaneous strain is small and the strain-gradient-induced polarization (flexoelectricity) is comparable to the biquadratic polarization. But even in such favorable material, the switchability of the twin wall polarization has not yet been established, nor for any other ferroelastic material [44].

A critical question thus emerges: Can the polarization of ferroelastic domain walls be switched by voltage? If so, how/when does it happen, given the competition between switchable (biquadratic) and non-switchable (flexoelectric) couplings? To answer these questions, we have calculated the effect of electric field on the ferroelastic domain walls of the orthorhombic Pnma phase of $CaTiO_3$. Using atomistic simulations, we determine that individual polar twin walls in $CaTiO_3$ can be switched by electric field. However, the switching is not ferroelectric-like, with a single hysteresis loop between two polar states, but antiferroelectric-like, with a double hysteresis between three: a flexoelectrically-dominated one at low fields and the two biquadratically-dominated polarities at high fields.

Our molecular dynamics (MD) simulations use empirical potentials developed by Pedone et al [45]. The MD simulation supercell consisted of $40 \times 14 \times 14$ formula units of CaTiO$_3$ (39200 atoms) with periodic boundary conditions applied in all three directions. NPT ensemble [46,47] (i.e., atomic number N, system pressure P and temperature T were held constant) were adopted with the system volumes allowed to change freely. We used the LAMMPS (Large-scale Atomic/Molecular Massively Parallel Simulator) codes [48], with visualizations by the OVTIO software [49]. For each simulation, a first 100000 steps (100ps) were performed to allow the structures to relax while another 100000 steps were followed to obtain the time averaged domain structures. To calculate the effect of an electric field (E), we add a force F=qE to each charged atom in the system.

We initially tested the phase transitions described by this potential. The phase transition sequence, along with respective tiltings of the TiO$_6$ octahedra, is $Cubic\left(Pm\bar{3}m, a^0 a^0 a^0\right) \rightarrow Tetragonal\left(I4/mcm, a^0 a^0 c^-\right) \rightarrow Orthorhombic\left(Pnma, a^- a^- c^+\right)$ (Figure 1(a-c)), and is in accordance with the experimental observations by Carpenter [50,51], Redfern [52] and Ali [53]. The cubic to tetragonal phase transition was found at 1330K, with the subsequent transition to the orthorhombic phase occurring at 1275K (Fig. 1(b)). These simulated transition temperatures are somewhat lower than the experimental values [45,54], but the phases themselves are correct and stable in the MD simulations, so the typical (110)-oriented ferroelastic DW of orthorhombic CaTiO$_3$ can be further investigated. The phase transition to the room-temperature orthorhombic phase involves rotations of the TiO$_6$ octahedra, resulting in a doubled pseudo-cubic cell along the *z* direction. In figure 1(d) we show the calculated structure of the odd and even layers along the z direction, used a bulk reference for the analysis of dipole moments inside the DWs.

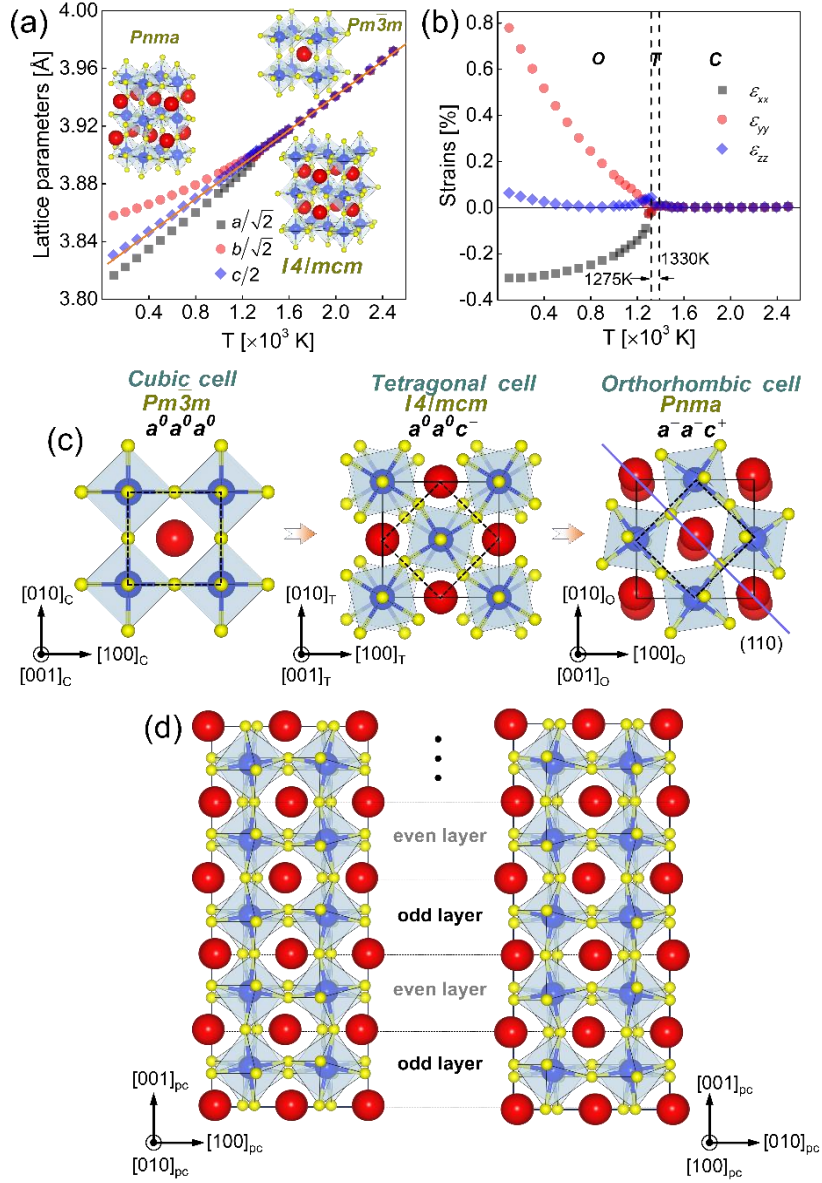

Figure 1 Phase transitions predicted by the empirical potential. (a) The temperature dependence of scaled lattice constants $a/\sqrt{2}$, $b/\sqrt{2}$ and $c/2$. (b) Strains $\varepsilon_{xx}$, $\varepsilon_{yy}$ and $\varepsilon_{zz}$ calculated with respect to the pseudocubic lattice parameters extrapolated from the high temperature cubic phase (red line in (a)). (c) crystalline structures of the cubic (space group, $Pm\bar{3}m$), tetragonal (space group, $I4/mcm$) and orthorhombic (space group, $Pnma$) phases. $a^0a^0a^0$, $a^0a^0c^-$ and $a^-a^-c^+$ denote the TiO$_6$ octahedra tilt models written in Glazer notation [55]. Black dashed rectangles in (c) indicate the reference cubic cell. (d) shows the odd and even perovskite layers along the $z$ direction of the pseudocubic cell with their respective TiO$_6$ octahedra rotations.

The structure of ferroelastic domain walls oriented along the (110) plane of orthorhombic CaTiO$_3$ at room temperature (T=300K) is shown in Fig. 2. Their geometry and orientation are schematically depicted in Fig. 2(a). The calculated twin angle in our simulation is c.a. 0.52°, somewhat smaller than the experimental value (~0.6°) in bulk [16], and larger than the value near the surface (~0.4°) [56], where surface relaxation by image forces tend to lower spontaneous strain [57,58]. Two ferroelastic DWs are included in the calculation (Fig. 2(b)). The polar vectors and octahedral tilts predicted by our simulations along *x*, *y* and *z* axes are shown in Fig. 2(c-h).

Our simulated polar textures at the DWs resemble quite well the experimental observation [16] (inset in Fig. 2(b)). The main polar components lay inside the twin wall plane along the *y* direction, which is the direction of maximum transverse strain gradient and thus maximum flexoelectricity. The main dipole components along the y direction for concave (valley-like) and convex (ridge-like) twin junctions are antiparallel (figure 2(c-d)), consistent the opposite signs of their strain gradients, and with the need to preserve global inversion symmetry. This pair-wise, gradient-driven antiparallel polarization of ferroelastic walls is fundamentally different from (and more general than) the depolarization-driven antipolar order of non-ferroelastic 180 degree walls in some hyperferroelectrics [59]. The maximum polarization density inside individual DWs in our simulation was calculated to be c.a. 2.4 $\mu$C/cm$^{-2}$, in the same range as reported by Van Aert et al [16].

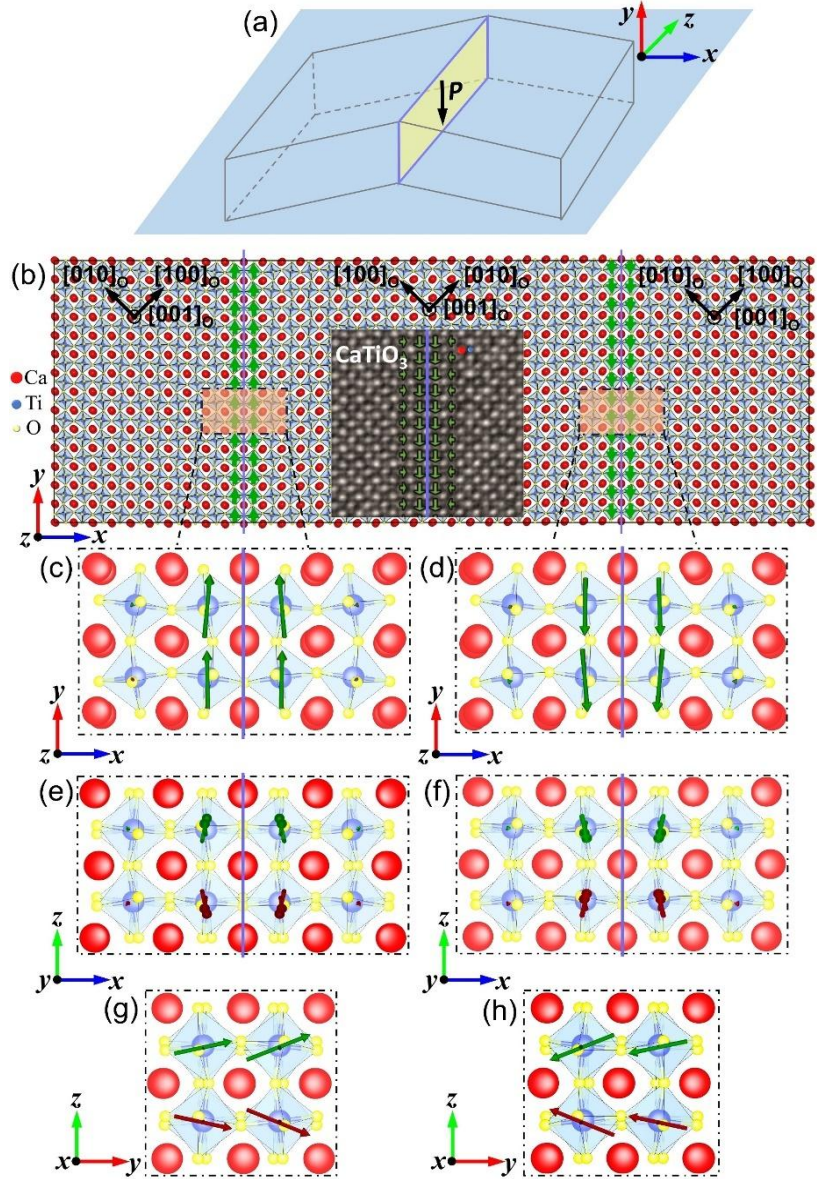

Figure 2 Structural and polar properties of (110) type DWs in orthorhombic $CaTiO_3$ at room temperature (T=300K). (a) Sketch of the geometry and orientation of the (110) DW in orthorhombic phase of $CaTiO_3$. (b) is the sandwich model consisting of two parallel DWs with antiparallel DW polar vectors (green arrows). The ferroelastic twin angle and maximum polarization inside the DW are c.a. 0.52° and 2.4 $\mu C/cm^{-2}$. Inset in (b) is the polar vectors inside DW obtained by transmission electron microscopy (TEM) [16]. Local details of DW polarizations inside *x-y*, *x-z* and *y-z* (twin wall plane) planes were shown in (c-d), (e-f) and (g-h). Green and red arrows are the polar vectors of odd and even layers corresponding to Fig. 1(d). Small polarization tilts exist inside *x-y* and *x-z* planes while relatively large tilts have been found inside the twin wall (*y-z*). The overall polarization vectors of nearby twin walls cancel out due to the conservation of global inversion centrosymmetry. The polar vector arrows are amplified by a factor of 150 for clarify.

We now look at the effect of external electric field on polarization. The first thing to notice is that, although the polarization of bulk $CaTiO_3$ is frustrated by octahedral tilts, a sufficiently large electric field can induce a phase transition to a polar phase. This is shown in figure 3, where the calculated bulk polarization and octahedral tilts around the *y* axis as a function of external electric field along the *y* axis are plotted. The antiferroelectric-like behaviour of bulk $CaTiO_3$ is noteworthy, and it already indicates that the energy balance between a tilt-dominated phase with suppressed polarization and a tilt-suppresed phase with emerging polarization can be changed by a sufficiently large electric field. With this in mind, let us examine the response of the domain walls.

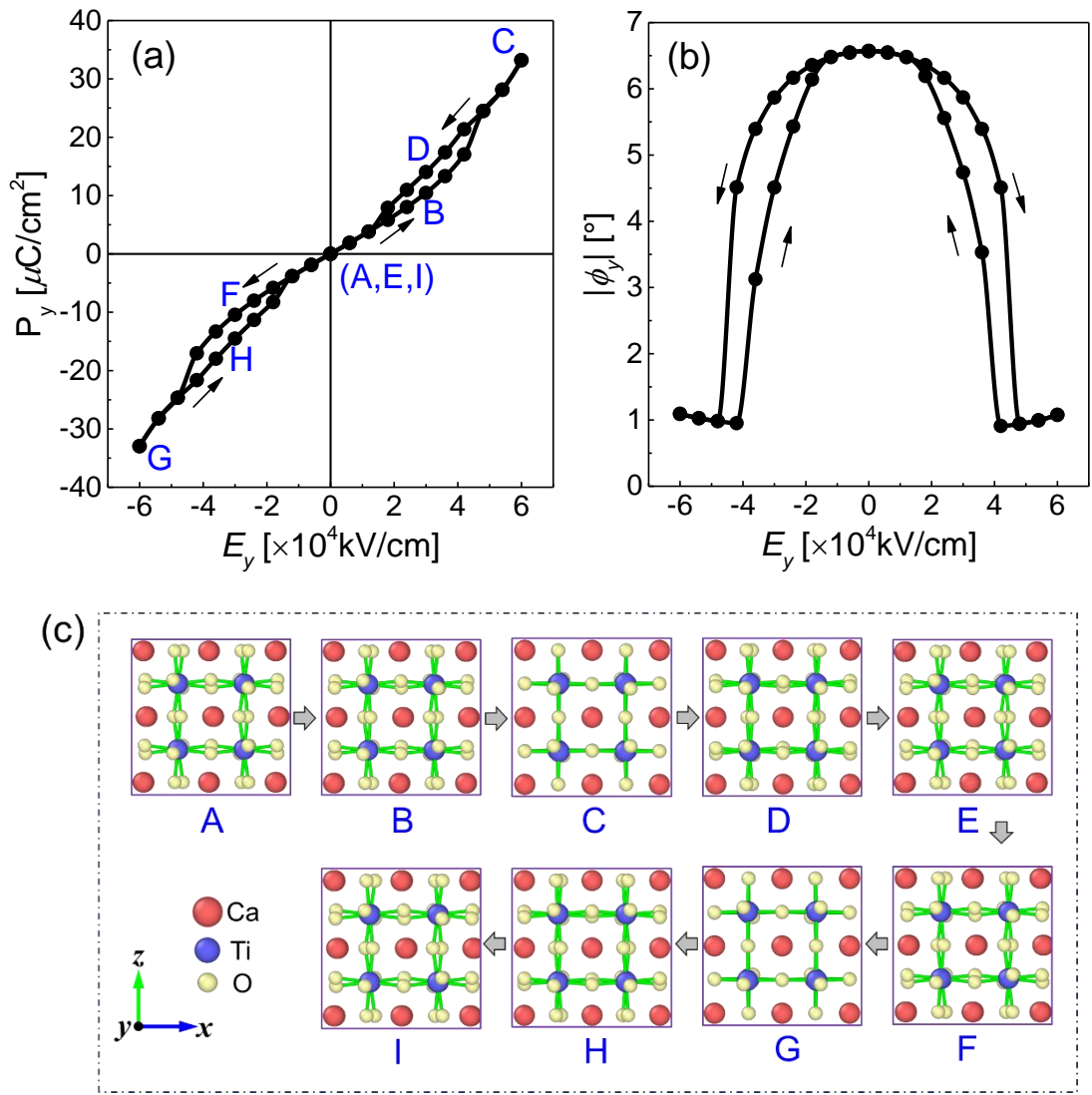

Figure 3 Polarization (a) and octahedral tilts (b) of $CaTiO_3$ around the *y* axies as a function of electric field along the *y* axis. A field-induced double-hysteretic transition to a phase with increased polarization, concomitant with a suppression of octahedral tilts, is observed. The local octahedral tilt structures at different electric field, as indicated by 'A-I' in (a) are shown in (c).

The structural response (twinning angle and polarization) of a $CaTiO_3$ supercell with two twin walls is shown in Figure 4. The polarization of the walls along the *y* direction, minus the substracted background bulk polarization ($P_y - P_y^{bulk}$), is plotted as function of electric field. Before applying electric field, ('A' in Fig. 4), the flexoelectric polarization is positive for one wall (red), negative for the other (blue), and zero on average, so there is no net remnant polarization. These polarizations are proportional to the strain gradient at the walls, and their antiparallel orientation is expected from symmetry. Applying an electric field along the positive y direction gradually polarizes the crystal structures of the domains and the domain walls. Concomitant with this increase in dielectric polarization, there is a decrease in spontaneous strain (signified by a decrease in the twinning angle), and a gradual broadening of the domain walls. Both these events decrease flexoelectricity –which, let us remember, is roughly equal to the spontaneous strain divided by the domain wall width. At some critical field (B) the weak flexoelectric polarization is overcome by the biquadratic term, and the polarization abruptly jumps ('B' in Fig. 4). Upon retrieving the electric field, the polar phase remains trapped in a metastable state ('C' in Fig. 4), and eventually reverts back to their initial state as the field is further lowered ('D' in Fig. 4). Applying the electric field along negative y direction, the DW dipoles behave analogously, thus generating an antiferroelectric-like double-hysteresis loop.

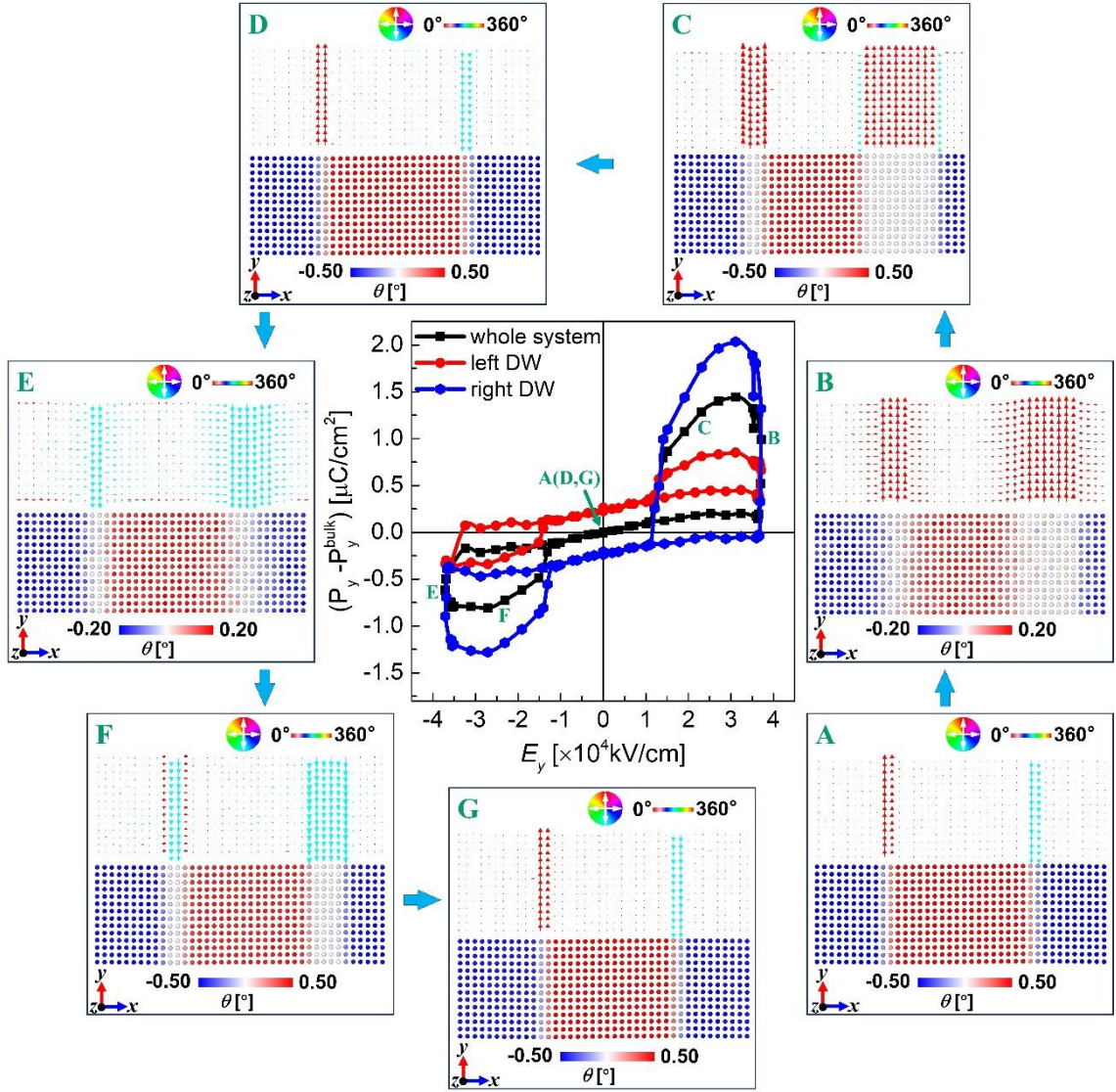

Figure 4 Antiferroelectric-like double hysteresis loops of polar twin walls in ferroelastic CaTiO$_3$ at T=300K. Insets 'A'-'G' indicate the strain and polar configurations under different electric fields. Ferroelastic twin structures in insets 'A'-'G' at different electric fields are described by the ferroelastic shear angle $\theta$. Polarization vectors in insets 'A'-'G' are coded by different colors according to their rotation angles (0-360°) around the $z$ axis. Polarization vectors are amplified by a factor of 1.5 for clarity.

Several conclusions can be drawn from these calculations. The first is that the internal polarization of ferroelastic domain walls is theoretically switchable. However, the switching is not ferroelectric-like but antiferroelectric-like. The lack of hysteresis at zero field means that ferroelastic DW polarization is not readily usable as a memory element, although the double hysteresis loops may find applications in miniaturized versions of antiferroelectric-like devices (e.g. local heating-cooling at the nanoscale via the electrocaloric effect [60-63]).

But the results also have fundamental repercussions at the macroscale of the crystal. As is visible in the calculations of the field-induced switched state inside the domain walls (panels B-C and E-F of figure 4), polar switching is concomitant with a broadening of the domain walls (such that the lower gradient reduces the unipolar flexoelectric bias). Further increasing the field over the maximum in the calculation of Figure 4 tips the entire crystal into the same bulk polarized state calculated in figure 3. The domain walls can thus act as the nucleation points for the field-polarized state in $CaTiO_3$, so that the ferroelastic walls of the non-polar phase broaden until they become the polar domains of the polarized phase.

Our results suggest that finely twinned crystals (e.g. thin films) can behave functionally as antiferroelectrics, with the domain walls acting both as polarizable structures in themselves, or as nuclei for a bulk phase transition to a polar state. Whether the field-induced transition from antiparallel polarization inside domain walls to parallel polarization in the bulk qualifies $CaTiO_3$ as a bona-fide antiferroelectric is open for debate, but in any case we note that this type behaviour could be very widespread, since the ferroelastic Pnma structure of $CaTiO_3$ is the most common symmetry among perovskites [55,64]. Among these are antiferromagnetic orthoferrites, which also raises the exciting prospect that their ferroelastic domain walls may behave as field-induced magnetoelectric multiferroics.

**ACKNOWLEDGMENTS**

Guangming Lu is grateful for the financial support by the National Natural Science Foundation of China (Grant No. 12304130), Shandong Provincial Natural Science Foundation (ZR2024QA146) and the Doctoral Starting Fund of Yantai University (Grant No. 1115-2222006). Gustau Catalan acknowledges financial support from the EU (FET-open grant N° 766726, project TSAR), from the Catalan government (grant number 2021 SGR 0129), and from the Agencia Estatal de Investigación (project numbers PID2023-148673NB-I00 and CEX2021-001214-S). Ekhard K. H. Salje is grateful to EPSRC (EP/P024904/1) and the EU's Horizon 2020 program under the Marie Sklodowska-Curie Grant (Grant No. 861153).

**NOTE**

After submission of this manuscript, Professor Ekhard Salje passed away. As his colleagues and co-authors, we take this opportunity to homage Ekhard, a gentleman, a scholar, and an endless source of inspiration.

# REFERENCES

[1] E. K. H. Salje, ChemPhysChem **11**, 940 (2010).

[2] G. F. Nataf, M. Guennou, J. M. Gregg, D. Meier, J. Hlinka, E. K. H. Salje, and J. Kreisel, Nature Reviews Physics **2**, 634 (2020).

[3] D. Meier and S. M. Selbach, Nature Reviews Materials **7**, 157 (2022).

[4] G. Catalan, J. Seidel, R. Ramesh, and J. F. Scott, Reviews of Modern Physics **84**, 119 (2012).

[5] T. Sluka, A. K. Tagantsev, P. Bednyakov, and N. Setter, Nature Communications **4**, 1808 (2013).

[6] K.-C. Kim *et al.*, Nature Communications **7**, 12449 (2016).

[7] D. M. Juraschek, Q. N. Meier, M. Trassin, S. E. Trolier-McKinstry, C. L. Degen, and N. A. Spaldin, Physical Review Letters **123**, 127601 (2019).

[8] D. M. Juraschek, M. Fechner, A. V. Balatsky, and N. A. Spaldin, Physical Review Materials **1**, 014401 (2017).

[9] T. Rojac *et al.*, Nature Materials **16**, 322 (2017).

[10] S. Farokhipoor and B. Noheda, Physical Review Letters **107**, 127601 (2011).

[11] J. Seidel *et al.*, Nature Materials **8**, 229 (2009).

[12] S. Y. Yang *et al.*, Nature Nanotechnology **5**, 143 (2010).

[13] A. Bhatnagar, A. Roy Chaudhuri, Y. Heon Kim, D. Hesse, and M. Alexe, Nature Communications **4**, 2835 (2013).

[14] S. Yun, K. Song, K. Chu, S.-Y. Hwang, G.-Y. Kim, J. Seo, C.-S. Woo, S.-Y. Choi, and C.-H. Yang, Nature Communications **11**, 4898 (2020).

[15] Y. Frenkel *et al.*, Nature Materials **16**, 1203 (2017).

[16] S. Van Aert, S. Turner, R. Delville, D. Schryvers, G. Van Tendeloo, and E. K. H. Salje, Advanced Materials **24**, 523 (2012).

[17] J. T. Eckstein, H. Yokota, N. Domingo, G. Catalan, O. Aktas, M. A. Carpenter, and E. K. H. Salje, Physical Review B **110**, 094107 (2024).

[18] G. Lu, S. Li, X. Ding, J. Sun, and E. K. H. Salje, Scientific Reports **9**, 15834 (2019).

[19] P. Zubko, G. Catalan, A. Buckley, P. R. L. Welche, and J. F. Scott, Physical Review Letters **99**, 167601 (2007).

[20] A. Alison and K. H. S. Ekhard, Journal of Physics: Condensed Matter **10**, L377 (1998).

[21] A. Aird, M. C. Domeneghetti, F. Mazzi, V. Tazzoli, and E. K. H. Salje, Journal of Physics: Condensed Matter **10**, L569 (1998).

[22]	D. V. Christensen *et al.*, Nature Physics **15**, 269 (2019).

[23]	Y. Frenkel, Y. Xie, H. Y. Hwang, and B. Kalisky, Journal of Superconductivity and Novel Magnetism **33**, 195 (2020).

[24]	G. Lu, S. Li, X. Ding, J. Sun, and E. K. H. Salje, npj Computational Materials **6**, 145 (2020).

[25]	G. Lu and E. K. H. Salje, APL Materials **12**, 061101 (2024).

[26]	E. K. H. Salje, Annual Review of Materials Research **42**, 265 (2012).

[27]	H. Huyan, L. Li, C. Addiego, W. Gao, and X. Pan, National Science Review **6**, 669 (2019).

[28]	Y.-P. Chiu *et al.*, Advanced Materials **23**, 1530 (2011).

[29]	B. Casals, A. Schiaffino, A. Casiraghi, S. J. Hämäläinen, D. López González, S. van Dijken, M. Stengel, and G. Herranz, Physical Review Letters **120**, 217601 (2018).

[30]	A. Schiaffino and M. Stengel, Physical Review Letters **119**, 137601 (2017).

[31]	L. Goncalves-Ferreira, S. A. T. Redfern, E. Artacho, and E. K. H. Salje, Physical Review Letters **101**, 097602 (2008).

[32]	G. Catalan, I. Lukyanchuk, A. Schilling, J. M. Gregg, and J. F. Scott, Journal of Materials Science **44**, 5307 (2009).

[33]	C.-L. Jia, S.-B. Mi, K. Urban, I. Vrejoiu, M. Alexe, and D. Hesse, Nature Materials **7**, 57 (2008).

[34]	P. V. Yudin, A. K. Tagantsev, E. A. Eliseev, A. N. Morozovska, and N. Setter, Physical Review B **86**, 134102 (2012).

[35]	O. Diéguez and M. Stengel, Physical Review X **12**, 031002 (2022).

[36]	Y. J. Wang, Y. L. Tang, Y. L. Zhu, Y. P. Feng, and X. L. Ma, Acta Materialia **191**, 158 (2020).

[37]	A. N. Morozovska, S. V. Kalinin, and E. A. Eliseev, in *Flexoelectricity in Solids* (WORLD SCIENTIFIC, 2015), pp. 311.

[38]	B. Houchmanzadeh, J. Lajzerowicz, and E. Salje, Phase Transitions **38**, 77 (1992).

[39]	B. Houchmandzadeh, J. Lajzerowicz, and E. Salje, Journal of Physics: Condensed Matter **3**, 5163 (1991).

[40]	M. Daraktchiev, G. Catalan, and J. F. Scott, Physical Review B **81**, 224118 (2010).

[41]	J. R. Kim *et al.*, Nature Communications **11**, 4944 (2020).

[42]	A. K. Tagantsev, E. Courtens, and L. Arzel, Physical Review B **64**, 224107 (2001).

[43]	A. N. Morozovska, E. A. Eliseev, M. D. Glinchuk, L.-Q. Chen, and V. Gopalan, Physical Review B **85**, 094107 (2012).

[44] G. Lu, S. Li, X. Ding, J. Sun, and E. K. H. Salje, Physical Review Materials **3**, 114405 (2019).

[45] A. Pedone, G. Malavasi, M. C. Menziani, A. N. Cormack, and U. Segre, The Journal of Physical Chemistry B **110**, 11780 (2006).

[46] W. G. Hoover, Physical Review A **31**, 1695 (1985).

[47] S. Nosé, The Journal of Chemical Physics **81**, 511 (1984).

[48] S. Plimpton, Journal of Computational Physics **117**, 1 (1995).

[49] A. Stukowski, Modelling and Simulation in Materials Science and Engineering **18**, 015012 (2010).

[50] M. A. Carpenter, A. I. Becerro, and F. Seifert, American Mineralogist **86**, 348 (2001).

[51] J. Manchado, F. J. Romero, M. C. Gallardo, J. del Cerro, T. W. Darling, P. A. Taylor, A. Buckley, and M. A. Carpenter, Journal of Physics: Condensed Matter **21**, 295903 (2009).

[52] A. T. R. Simon, Journal of Physics: Condensed Matter **8**, 8267 (1996).

[53] R. Ali and M. Yashima, Journal of Solid State Chemistry **178**, 2867 (2005).

[54] A. Pedone, The Journal of Physical Chemistry C **113**, 20773 (2009).

[55] A. Glazer, Acta Crystallographica Section B **28**, 3384 (1972).

[56] G. Magagnin, C. Lubin, M. Escher, N. Weber, L. Tortech, and N. Barrett, Physical Review Letters **132**, 056201 (2024).

[57] B. Gurrutxaga-Lerma, D. S. Balint, D. Dini, and A. P. Sutton, Proceedings of the Royal Society A: Mathematical, Physical and Engineering Sciences **471**, 20150433 (2015).

[58] G. Lu, X. Ding, J. Sun, and E. K. H. Salje, Physical Review B **106**, 144105 (2022).

[59] M. Conroy *et al.*, Advanced Materials **36**, 2405150 (2024).

[60] B. Peng, H. Fan, and Q. Zhang, Advanced Functional Materials **23**, 2987 (2013).

[61] B. Peng, Q. Zhang, Y. Lyu, L. Liu, X. Lou, C. Shaw, H. Huang, and Z. Wang, Nano Energy **47**, 285 (2018).

[62] P. Vales-Castro *et al.*, Physical Review B **103**, 054112 (2021).

[63] M. Li, S. Han, Y. Liu, J. Luo, M. Hong, and Z. Sun, Journal of the American Chemical Society **142**, 20744 (2020).

[64] M. A. Peña and J. L. G. Fierro, Chemical Reviews **101**, 1981 (2001).